\documentclass{article}
\usepackage{spconf,amsmath,graphicx}
\usepackage{float}
\usepackage{bbding}
\usepackage{color}
\usepackage{enumitem}
\usepackage{multirow}
\usepackage{etoolbox,siunitx}
\robustify\bfseries
\usepackage{booktabs}
\usepackage{balance}
\usepackage{amssymb}
\usepackage{bm}
\usepackage{arydshln}
\usepackage{caption}
\usepackage{cite}
\usepackage{hyperref}
\usepackage{circledsteps}
\usepackage{siunitx}

\usepackage{amssymb}
\usepackage{pifont}

\def\RR{{\mathbb R}}

\usepackage{caption}
\makeatletter
\renewcommand\normalsize{%
   \@setfontsize\normalsize{9pt}{10.5pt}} 
\makeatother

\let\OLDthebibliography\thebibliography
\renewcommand\thebibliography[1]{
  \OLDthebibliography{#1}
  \setlength{\parskip}{0.5pt}
  \setlength{\itemsep}{1pt plus 0.3ex}
}
\usepackage{flushend}
\title{MC-LExt: Multi-Channel Target Speaker Extraction with Onset-Prompted Speaker Conditioning Mechanism}

\name{Tongtao Ling$^{1}$, Shulin He$^{1}$, Pengjie Shen$^{2}$, and Zhong-Qiu Wang$^{1}$
}
\address{
$^{1}$Southern University of Science and Technology, Shenzhen, China\\
$^{2}$Inner Mongolia University, Hohhot, China\\
{\small\texttt{lingtt2025@mail.sustech.edu.cn, 
wang.zhongqiu41@gmail.com
}}
}
\begin{document}
\ninept
\maketitle
\begin{abstract}
Multi-channel target speaker extraction (MC-TSE) aims to extract a target speaker's voice from multi-speaker signals captured by multiple microphones.
Existing methods often rely on auxiliary clues such as direction-of-arrival (DOA) or speaker embeddings.
However, DOA-based approaches depend on explicit direction estimation and are sensitive to microphone array geometry, while methods based on speaker embeddings model speaker identity in an implicit manner and may degrade in noisy-reverberant conditions.
To address these limitations, we propose \textit{multi-channel listen to extract} (MC-LExt), a simple but highly-effective framework for MC-TSE.
Our key idea is to prepend a short enrollment utterance of the target speaker to each channel of the multi-channel mixture, providing an onset-prompted conditioning signal that can guide TSE.
This design allows the deep neural network (DNN) to learn spatial and speaker identity cues jointly in a fully end-to-end manner.
Experiments on noisy-reverberant benchmarks, including WHAMR! and MC-Libri2Mix, demonstrate the effectiveness of MC-TSE.
%
\end{abstract}
\section{Introduction}

Target speaker extraction (TSE) aims to separate the speech of a targeted speaker from a mixture speech containing multiple speakers given an auxiliary clue.
It has broad applications in real-world scenarios such as virtual assistants, hearing aids, and smartphones \cite{zmolikova2023neural}.
While recent studies~\cite{quan2024spatialnet, wang2023tf} have achieved notable progress in monaural TSE under anechoic scenarios, many existing methods struggle in noisy-reverberant, multi-microphone settings \cite{zorilua2021investigation}, where complex acoustic interference and spatial ambiguity can cause significant performance degradation. 

Existing multi-channel TSE (MC-TSE) approaches generally fall into two categories: methods that rely on pre-extracted speaker embeddings to condition the network, and methods that exploit direction-of-arrival (DOA) cues to utilize spatial information~\cite{zhang2025doa}. For speaker-embedding-based methods, a monaural enrollment utterance is typically recorded and a speaker model is utilized to obtain a fixed-length speaker embedding, which serves as a global identity cue to guide the extraction network~\cite{vzmolikova2019speakerbeam,zhang2025multi}. In addition, the speaker embedding can be integrated into the intermediate layers of the extraction network, enabling more fine-grained guidance~\cite{he2024hierarchical,wang2018deep,delcroix2020improving,wang2024wesep,perez2018film}. 
However, speaker embedding only provides a fixed-length vector of the target speaker and lacks frame-level details that are helpful for precise TSE.
Differently, DOA-based methods aim to model the acoustic scene and localize the target speaker within the mixture.
By leveraging the spatial cues, the extraction network can separate overlapping speakers based on their spatial positions in noisy-reverberant environments~\cite{choi2025multichannel,li2019direction,han2021multi}.
Nevertheless, these approaches usually require explicit DOA estimation or carefully designed beamforming techniques~\cite{souden2009optimal,gu2019neural,elminshawi2023beamformer}, which can be sensitive to noise, reverberation, and microphone array geometry, and they become less effective when the speakers are spatially close to each other.

On the other hand, both of the multi-channel methods often rely on complex pre-processing pipelines and may not generalize well across diverse acoustic environments, motivating the need for a simpler and more robust paradigm.
Recently, the \textit{listen to extract} (LExt) technique~\cite{shen2025listen} introduces an extremely-simple while highly-powerful method for monaural TSE.
It first pre-pends an enrollment utterance to each training mixture, and then trains a DNN on the resulting mixtures to predict the target speech.
It shows that the pre-pended enrollment utterance can serve as a very effective prompt for the TSE network to identify the target speaker and extract the target speech.
In this context, this paper aims to design a prompting-based multi-channel TSE system that avoids explicit spatial modeling while obtaining strong TSE performance.

To this end, we extend LExt to the multi-channel setting, proposing \textit{multi-channel LExt} (MC-LExt).
MC-LExt prepends a short enrollment utterance to each mixture channel, constructing a speaker conditioning signal that triggers the extraction process.
Considering that the prepending mechanism results in a longer signal to process, to reduce the computation cost, we design a lightweight downsampler to reduce the computation spent on the enrollment speech.
Furthermore, we propose to integrate conventional speaker embedding based conditioning mechanisms into MC-LExt to ensure that sufficient speaker information is retained to guide the model.
Experiments on the WHAMR!~\cite{Maciejewski2020} and MC-Libri2Mix~\cite{ge2022spex} datasets demonstrate that MC-LExt achieves strong TSE performance
in noisy-reverberant conditions.
The contributions of this paper can be summarized as follows:
\begin{itemize}[leftmargin=*,noitemsep,topsep=0pt]
\item We propose MC-LExt, a simple but highly-effective framework for MC-TSE.
It eliminates the need for explicit spatial parameter estimation, enabling fully end-to-end training for TSE.
\item We introduce a computationally-efficient down-sampling module to reduce the computation spent for the enrollment utterance.
\item We integrate speaker embedding based conditioning into MC-LExt to provide sufficient speaker identity information for TSE.
\end{itemize}


\section{Review of Monaural LExt}\label{review_LExt}

LExt~\cite{shen2025listen} proposes a simple and effective approach for monaural TSE.
It prepends an enrollment utterance to each training mixture and trains DNNs to extract the target speaker based on the prepended mixtures.
The rationale is that prepending the enrollment utterance can create an earliest speech onset for the target speaker and such an onset can prompt DNNs to extract the target speaker based on the prepended mixture.
This onset-prompting strategy enables the model to identify the target speaker without relying on explicit speaker embeddings, getting rid of complicated DNN modules fusing speaker and mixture embeddings.
Despite its success in monaural settings, LExt has not been examined in multi-channel scenarios, where spatial cues may interact with the onset prompt in non-trivial ways.
Moreover, in LExt, longer enrollment utterance prepended to the mixture increases the length of the signal to process and hence requires more computation.
This could limit the practical deployment of LExt.
In this work, we extend LExt to multi-channel TSE by designing efficient prompting strategies that reduce the computation spent on enrollment speech, while incorporating speaker embedding based conditioning into MC-LExt, aiming to balance the efficiency and extraction performance.

\section{MC-LExt}\label{proposedalgorithm}

\subsection{Multi-Channel Listen to Extract (MC-LExt)}

\begin{figure}
    \centering
    \includegraphics[width=1.0\linewidth]{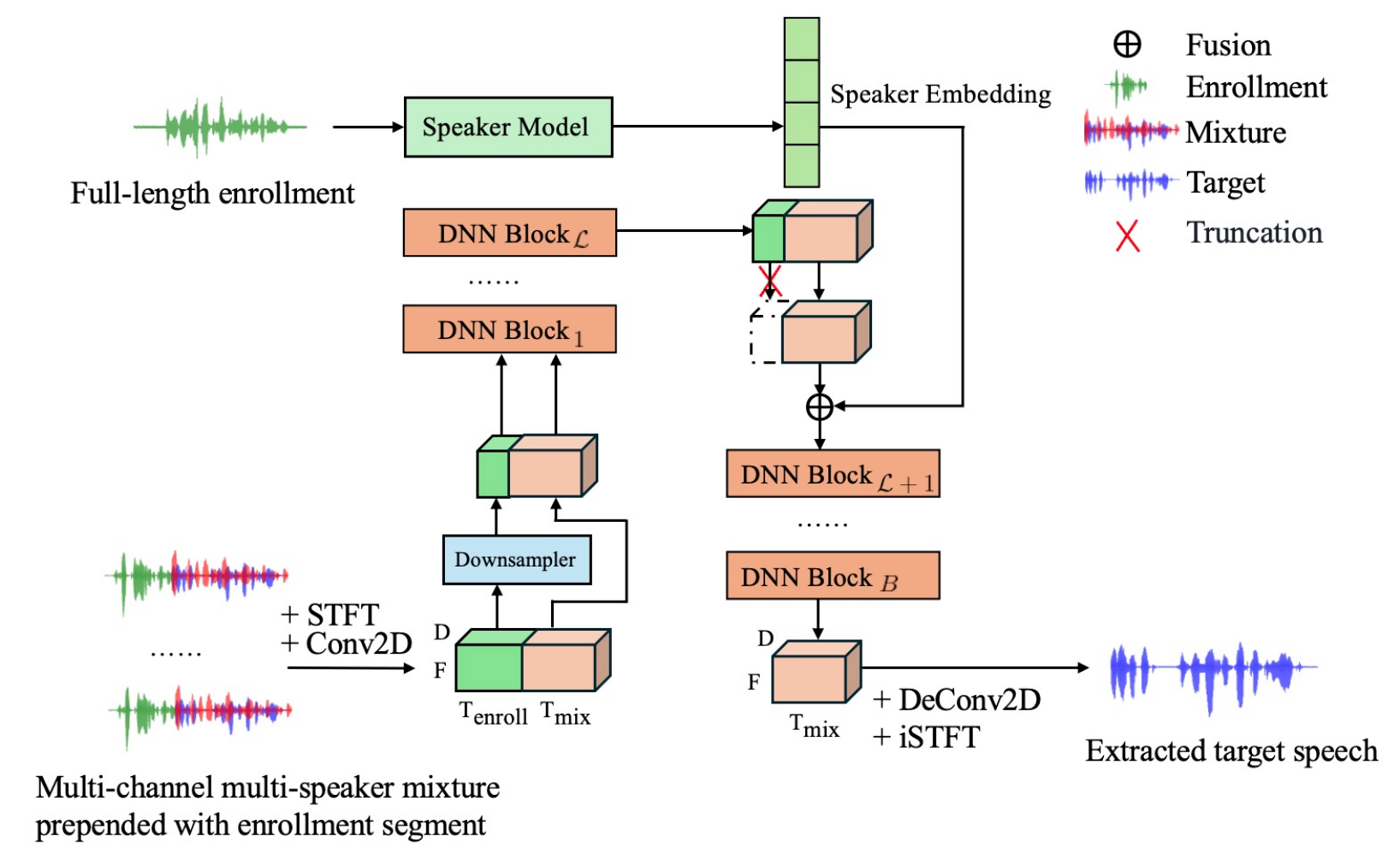}
    \vspace{-0.6cm}
    \caption{Overview of MC-LExt. Best viewed in color.}
    \vspace{-0.5cm}
    \label{fig:model}
\end{figure}

Fig.~\ref{fig:model} illustrates the proposed MC-LExt.
Given a multi-channel mixture $y \in \mathbb{R}^{C \times N}$ and the corresponding clean target $s \in \mathbb{R}^{N}$, the goal is to reconstruct $s$ from the mixture $y$.
Following monaural LExt \cite{shen2025listen}, MC-LExt prepends an enrollment utterance $e \in \mathbb{R}^{E}$ to each channel of the mixture, forming the input to the DNN for MC-TSE.
In the time domain, this procedure can be denoted as
\begin{align}
\widetilde{y} = [\, \widetilde{y}_1, \widetilde{y}_2, \dots, \widetilde{y}_C \,]^\top \in \mathbb{R}^{C \times (E+N)},\,\text{with}\,\,\widetilde{y}_c = [\, e; y_c \,],
\end{align}
where $E$ and $N$ are respectively the number of time-domain samples of the enrollment and mixture, and $C$ is the number of microphone channels.
Based on the prepended mixture, $\widetilde{y}$, MC-LExt trains a DNN to predict the target speech $s$ via supervised learning.
The prepending strategy introduces a unified onset-like prompt to all microphones simultaneously, enabling the DNN to condition its TSE process on the temporal and spectral characteristics afforded by the enrollment utterance.

In detail, as shown in Fig.~\ref{fig:model}, after utterance-level concatenation, we transform
the fused signal $\widetilde{y}$ into T-F domain via short-time Fourier transform (STFT).
We further compute the real and imaginary (RI) components and append the magnitude spectrum as an additional feature map, resulting in a stacked representation:
\begin{align}
    \mathbf{X}_{\text{Spec}} &= \mathrm{STFT}(\widetilde{y}) \in \mathbb{C}^{ C \times T \times F}, \\
    \mathbf{X}_{\text{RI+Mag}} &= \big[\mathcal{R}(\mathbf{X}_{\text{Spec}}), \mathcal{I}(\mathbf{X}_{\text{Spec}}), |\mathbf{X_{\text{Spec}}}|\big] \in \mathbb{R}^{ (2C+1) \times T \times F},
\end{align}
where $\mathcal{R}(\cdot)$, $\mathcal{I}(\cdot)$ and $|\cdot|$ respectively extract the real component, imaginary component and magnitude of a complex spectrogram, $F$ denotes the number of frequency bins, and $T=T_\text{enroll} + T_\text{mix}$ denotes the number of time frames, with $T_\text{enroll}$ and $T_\text{mix}$ representing the number of frames of the enrollment segment and the original mixture.
We then apply a Conv2D layer over the concatenated input $\mathbf{X}_{\text{RI+Mag}}$ to project it into a higher-dimensional embedding:
\begin{align}
    \mathbf{H} = \mathrm{Conv2D}(\mathbf{X}_{\text{RI+Mag}}) \in \mathbb{R}^{ D \times T \times F},
\end{align}
where $D$ denotes the embedding dimension of each T-F unit.
This high-dimensional representation is then processed by a stack of DNN blocks to extract the target speech.
To reduce computation, we propose to \Circled{\footnotesize 1} downsample the embedding of the enrollment segment along time and \Circled{\footnotesize 2} only pass it through a subset of the DNN blocks, whereas the mixture segment is fully processed by all the DNN blocks.
Additionally, a fixed-length speaker embedding is computed from the full-length enrollment and we propose to \Circled{\footnotesize 3} use it to condition the MC-LExt model, helping the MC-LExt model which already leverages onset-based conditioning to better extract the target speaker.
Finally, the DNN blocks produce a high-dimensional representation $\mathbf{\widetilde{Z}}_B^{\text{mix}}\in \RR^{D\times T_\text{mix}\times F}$, and a 2D transposed convolution (DeConv2D) is applied to project it back to the complex spectrogram space:
\begin{equation}
    \mathbf{S} = \mathrm{DeConv2D}(\mathbf{\widetilde{Z}}_B^{\text{mix}})  \in \mathbb{R}^{ 2  \times T_{\text{mix}} \times F },
\end{equation}
where the first dimension corresponds to the RI components of the estimated spectrogram, and $B$ is the number of DNN blocks.
Then, iSTFT is applied to estimate the target speech in the time domain:
\begin{equation}
    \hat{s} = \mathrm{iSTFT(\mathbf{S})} \in \mathbb{R}^{N}.
\end{equation}
In default, the DNN is trained using the scale-invariant signal-to-distortion ratio (SI-SDR) loss \cite{le2019sdr}.

In the rest of this section, we describe the proposed techniques marked using \Circled{\footnotesize 1}, \Circled{\footnotesize 2} and \Circled{\footnotesize 3} in the previous paragraph.

\subsection{Downsampling Enrollment Utterance}

The computation of MC-LExt grows linearly with the prepended enrollment utterance.
To reduce the computation while preserving essential spectro-temporal information in the enrollment utterance, we apply a \texttt{downsampler} module consisting of a stack of Conv2D blocks to reduce the time steps of the embedding of the enrollment utterance.
Each convolution block consists of a GroupNorm, ReLU activation, and a Conv2D layer with a kernel size of $3 \times 3$ and a stride of $2 \times 1$.
This effectively halves the temporal resolution at each block while maintaining the frequency dimension, i.e., 
\begin{equation}
    \text{Downsampler} = \big[\text{Conv2D, ReLU, GroupNorm} \big]_{\times G},  
\end{equation}
where $G = \log_2(\lfloor\frac{E}{E^\prime}\rfloor)$ and $E^\prime$ is the length of compressed enrollment speech. We then use the downsampler to compress the enrollment segment:
\begin{equation}
    \mathbf{U} = \big[\text{Downsampler}(\mathbf{H}_\text{enroll}); \mathbf{H}_\text{mix}\big] \in \mathbb{R}^{ D \times T^{\prime} \times F},\label{U_description}
\end{equation}
where $\mathbf{H}_\text{enroll}$ and $\mathbf{H}_\text{mix}$ respectively denote the embeddings of enrollment segment and mixture segment. 
After that, $\mathbf{U}$ is fed into a stack of DNN blocks (e.g., TF-GridNet blocks~\cite{wang2023tf}) for training. 

\subsection{Selective Blocks for Enrollment Segment}

To further reduce computation, we propose to use a subset of DNN blocks to process the entire prepended signal (consisting of the enrollment segment and mixture segment), while the remaining blocks operate solely on the mixture segment.
Specifically, let the input be $\mathbf{Z}_0 = \mathbf{U}$ in Eq. (\ref{U_description}), the output of the first $\mathcal{L}$-th blocks are computed as
\begin{equation}
    \mathbf{Z}_i = \text{Block}_i(\mathbf{Z}_{i-1}), \quad 1 \leq i \leq \mathcal{L}.
\end{equation}
where $\mathcal{L}$ is a tunable hyper-parameter. After the first $\mathcal{L}$ blocks, to reduce computation we discard the enrollment segment,
and retain only the last $T_{\text{mix}}$ frames corresponding to the mixture segment, i.e., $\mathbf{Z}_\mathcal{L}^{\text{mix}} = \mathbf{Z}_\mathcal{L}[:,T^{\prime}-T_{\text{mix}}:,:]$. The subsequent blocks process the intermediate representations in the following way:
\begin{equation} \label{eq:forward}
    \mathbf{Z}_j^{\text{mix}} = \text{Block}_j(\mathbf{Z}_{j-1}^{\text{mix}}), \quad \mathcal{L}+1\leq j \leq B. \\
\end{equation}



\subsection{Speaker Embedding Fusion}

The intermediate mixture representation $\mathbf{Z}_j^{\text{mix}}$ can be further fused by incorporating speaker identity information. In real-world scenarios, enrollment utterances are often captured via microphone under less constrained conditions and can be very long (e.g., $30$ seconds or more). In MC-LExt, directly prepending such lengthy enrollments for TSE is computationally expensive and inefficient.
Therefore, we maintain the speaker embedding conditioning mechanism and use a speaker model (e.g., ECAPA-TDNN~\cite{desplanques2020ecapa}) to obtain fixed-length embeddings from full-length enrollment utterances.
It can be fused with $\mathbf{Z}_j^{\text{mix}}$ through various fusion methods, such as concatenation~\cite{wang2018deep}, addition~\cite{wang2024wesep}, multiplication~\cite{delcroix2020improving} and FiLM~\cite{perez2018film}).
The fusion process is defined as follows:
\begin{equation}
    \mathbf{\widetilde{Z}}_j^{\text{mix}} = \texttt{Fusion}(\mathbf{Z}_j^{\text{mix}},v) \in \mathbb{R}^{D \times T_{\text{mix}} \times F},
\end{equation}
where $v \in \mathbb{R}^{K}$ is a fixed-length speaker embedding.
When using this method, we replace $\mathbf{Z}_j^{\text{mix}}$ in Eq.~(\ref{eq:forward}) with $\mathbf{\widetilde{Z}}_j^{\text{mix}}$.



\subsection{Loss Functions for Negative Pairs}

MC-LExt is trained in an end-to-end manner to optimize the quality of the reconstructed target speech using the SI-SDR loss. To improve model robustness, we further consider the situation where the enrollment speaker is not present in the mixture, in which case the model should output silence. We adopt a contrastive training objective: for positive pairs, where the enrollment speaker is present in the mixture, the model is trained to reconstruct target speaker speech; for negative pairs, where the enrollment speaker is absent in the mixture, the model is instead encouraged to output silence. The standard SI-SDR loss is undefined for a silent target signal, as the scaling factor $\alpha = \frac{\langle \hat{s}, s \rangle}{|s|^2}$ becomes undefined due to a division by $0$ when the target signal $s$ is all zeros. Using the standard SI-SDR loss would thus lead to numerical instability and training failure. Following previous work~\cite{wisdom2021s}, we use log-MSE loss to handle negative pairs. Given the mixture signal $y$, target signal $s$, and estimated output $\hat{s}$, the log-MSE loss is defined as
\begin{equation}
\text{LOG-MSE}(s,\hat{s},y) = 
\begin{cases}
10 \log_{10}(||s-\hat{s}||^2+\tau||s||^2), & s \neq \mathbf{0} \\
10 \log_{10}(||\hat{s}||^2+\tau||y||^2), & s = \mathbf{0}
\end{cases} \nonumber
\end{equation}
where $\tau=10^{-\text{SNR}_{\text{max}}/10}$ and $\text{SNR}_{\text{max}}$ is set to $30$ dB.

\section{Experimental Setup}\label{setup}


We conduct experiments on two datasets:
1) \textbf{WHAMR!}~\cite{Maciejewski2020}:
Each mixture comprises two concurrent speech sources from the original WSJ0-2mix dataset, combined with a noise signal sampled from WHAM!~\cite{Wichern2019WHAM} and convolved with room impulse responses (RIRs) simulated using the Pyroomacoustics toolkit~\cite{scheibler2018pyroomacoustics}. The dataset contains $20,000$ training, $5,000$ validation, and $3,000$ test utterances of two-speaker mixed speech.
2) \textbf{MC-Libri2Mix}~\cite{ge2022spex}: a multi-channel extension of the original Libri2Mix dataset~\cite{cosentino2020librimix}, containing $2$-speaker mixtures recorded with $4$ microphones.
It comprises $63,528$ training, $1,172$ validation, and $3,000$ test utterances of two-speaker mixed speech.
For both datasets, we use $8$ kHz sampling rate and the \textit{min} version.
For WHAMR!, we adopt the $2$-channel version, and, for MC-Libri2Mix, the $4$-channel version.
Each speaker enrollment utterance is randomly selected from the original WSJ0~\cite{hershey2016deep} and LibriSpeech~\cite{panayotov2015librispeech} corpus.
For evaluation, the enrollment utterances are chosen following the same setting as in prior works~\cite{shen2025listen,ge2022spex}.

\begin{table}[!tp]
\scriptsize
    \centering
    \sisetup{table-format=2.2,round-mode=places,round-precision=2,table-number-alignment = center,detect-weight=true,detect-inline-weight=math}
    \caption{
    Results on WHAMR!.
    ``DS'' means whether using downsampler and ``SE'' means whether speaker embedding.
    }
    \vspace{-0.3cm}
    \setlength{\tabcolsep}{2pt}
    \begin{tabular}{
    cccccc
    S[table-format=2.1,round-precision=1]
    S[table-format=2.1,round-precision=1]
    }
    \toprule
    System & DNN arch. & $C$ & DS & SE & Loss &  {SDRi (dB)} & {SI-SDRi (dB)} \\
    \midrule
    LExt~\cite{shen2025listen} & TFGridNetV1 & 1 & \XSolidBrush & \XSolidBrush & SI-SDR  & 15.5 & 17.1 \\
    \hline
    Vanilla TSE & TFGridNetV1 & 2 & \XSolidBrush & \CheckmarkBold & SI-SDR & 15.9 & 17.4 \\
    MC-LExt & TFGridNetV1 & 2 & \XSolidBrush & \XSolidBrush & SI-SDR & 17.1 & 18.6 \\
    MC-LExt & TFGridNetV1 & 2 & \CheckmarkBold & \XSolidBrush & SI-SDR & 17.2 & 18.8 \\
    MC-LExt & TFGridNetV1 & 2 & \XSolidBrush & \CheckmarkBold & SI-SDR & 17.4 & 18.9 \\
    MC-LExt & TFGridNetV1 & 2 & \CheckmarkBold & \CheckmarkBold & SI-SDR & \bfseries 17.6 & \bfseries 19.1 \\
    MC-LExt & TFGridNetV1 & 2 & \CheckmarkBold & \CheckmarkBold & LOG-MSE & 17.3 & 18.8 \\
    \bottomrule
    \end{tabular}
    \label{tab:ablation}
\footnotesize\textit{Notes}: Using $8$-second enrollment utterance when downsampler is available.
\end{table}

\begin{table}[!tp]
        \vspace{-0.3cm}
\scriptsize
    \centering
    \sisetup{table-format=2.2,round-mode=places,round-precision=2,table-number-alignment = center,detect-weight=true,detect-inline-weight=math}
    \caption{Energy suppression ratio on negative pairs for Vanilla TSE and MC-LExt based on WHAMR! test set.}
            \vspace{-0.3cm}
    \label{tab:energy}
    \begin{tabular}{
    l
    S[table-format=2.1,round-precision=1]
    }
    \toprule
    System & {Energy suppression ratio (dB)} \\
    \midrule
    Vanilla TSE & 45.7 \\
    MC-LExt     & \bfseries 61.0 \\
    \bottomrule
    \end{tabular}
\end{table}

\begin{table}[!tp]
        \vspace{-0.2cm}
\scriptsize
\centering
\sisetup{table-format=2.2,round-mode=places,round-precision=2,table-number-alignment = center,detect-weight=true,detect-inline-weight=math}
\caption{Results of various speaker embedding fusion types (without downsampler and with SI-SDR loss).}
\vspace{-0.3cm}
\begin{tabular}{
ccc
S[table-format=2.1,round-precision=1]
S[table-format=2.1,round-precision=1]
}
\toprule
System & Fusion type & DNN arch. & {SDRi (dB)} & {SI-SDRi (dB)} \\
\midrule
MC-LExt & Concatenation & TFGridNetV1 & 17.4 & 18.9  \\
MC-LExt & Addition & TFGridNetV1  & 17.2 & 18.6  \\
MC-LExt & Multiplication & TFGridNetV1 & \bfseries 17.6 & \bfseries 19.1  \\
MC-LExt & FiLM & TFGridNetV1 & 17.4 & 18.8 \\
\bottomrule
\end{tabular}
    \vspace{-0.2cm}
\label{tab:fusion}
\end{table}

\begin{table}[!t]
\scriptsize
\centering
\sisetup{table-format=2.2,round-mode=places,round-precision=2,table-number-alignment = center,detect-weight=true,detect-inline-weight=math}
\caption{Results of using various number of enrollment forward blocks on $2$-channel WHAMR! dataset (without downsampler and speaker embedding, and with SI-SDR loss). GMAC/S is computed based on a $4$-second mixture and $4$-second enrollment speech.}
\vspace{-0.3cm}
\begin{tabular}{
cc
S[table-format=3.1,round-precision=1]
S[table-format=2.1,round-precision=1]
S[table-format=2.1,round-precision=1]
}
\toprule
System & DNN arch. & {GMAC/s} & {SDRi (dB)} & {SI-SDRi (dB)}\\
\midrule
MC-LExt ($\mathcal{L}$=4) & TFGridNetV1 & 316.73 & 17.1 & 18.6  \\
MC-LExt ($\mathcal{L}$=3) & TFGridNetV1 & 277.31 & \bfseries 17.7 & \bfseries 19.2 \\
MC-LExt ($\mathcal{L}$=2) & TFGridNetV1 & 237.89 & 17.4 & 19.0  \\
MC-LExt ($\mathcal{L}$=1) & TFGridNetV1 & 198.48 & 17.4 & 18.9  \\
\bottomrule
\end{tabular}
    \vspace{-0.2cm}
\label{tab:forward_blocks}
\end{table}


For STFT, we use $16$ ms window size, $8$ ms hop size, and the square-root Hanning analysis window. Each training sample consists of a $4$-second segment of enrollment utterance and a $4$-second segment of mixture.
When a downsampler is applied, an $8$-second enrollment segment is used and is downsampled by $50\%$.
We use TFGridNet~\cite{wang2023tf} with two configurations as the DNN architectures.
For \textbf{TFGridNetV1}, we use $B=4$ TF-GridNet blocks, each configured with the following hyper-parameters: $D = 128, I=1, J=1, H=200, E=16$, and $L=4$, following the notations of TF-GridNet \cite{wang2023tf}.
For \textbf{TFGridNetV2}, we set them to $B = 6, D = 128, I=1, J=1, H=256, E=16$, and $L=4$.
The smaller V1 model is used to verify the effectiveness of each proposed component, while the larger V2 model is used for comparison with state-of-the-art systems.
We use ECAPA-TDNN~\cite{desplanques2020ecapa} to extract $192$-dimensional speaker embeddings.
MC-LExt is trained for up to $100$ epochs with ADAM~\cite{kingma2014adam}.
The learning rate starts from $5\times10^{-4}$ and is halved after $2$ stagnant validation epochs.
Performance is measured using SI-SDR improvement (SI-SDRi) and SDR improvement (SDRi).




\begin{table}[!tp]
\scriptsize
    \centering
    \sisetup{table-format=2.2,round-mode=places,round-precision=2,table-number-alignment = center,detect-weight=true,detect-inline-weight=math}
    \caption{Results on $4$-channel MC-Libri2Mix dataset. $^{\ast}$ denotes results reproduced by us.}
    \vspace{-0.2cm}
    \setlength{\tabcolsep}{3pt}
    \begin{tabular}{{
    l
    S[table-format=2.1,round-precision=1]
    S[table-format=2.1,round-precision=1]
    }}
    \toprule
    System & {SDRi (dB)} & {SI-SDRi (dB)} \\
    \midrule
    Mask-based MVDR Beamforming~\cite{erdogan2016improved} & 7.6  & 6.2  \\
    Pretrained Speaker Localizer~\cite{ge2022spex} & 7.0  & 5.7 \\
    L-SpEx w/o E2E Train~\cite{ge2022spex} & 9.0  & 7.2 \\
    L-SpEx~\cite{ge2022spex} & 9.2 & 7.4 \\
    HSRL-TSE$^{\ast}$~\cite{he2025enhancing} & 8.6 & 8.4 \\
    \midrule
    MC-LExt (TFGridNetV2) \& SI-SDR & \textbf{18.6}  & \textbf{16.3}  \\
    MC-LExt (TFGridNetV2) \& LOG-MSE & 16.5  & 15.6  \\
    \bottomrule
    \end{tabular}
    \label{tab:libri2mix}
\end{table}
\begin{table}[!tp]
\scriptsize
    \vspace{-0.2cm}
    \centering
    \sisetup{table-format=2.2,round-mode=places,round-precision=2,table-number-alignment = center,detect-weight=true,detect-inline-weight=math}
    \caption{Comparison of MC-LExt with other systems based on $2$-channel WHAMR! dataset. $^{\ast}$ denotes results reproduced by us.}
        \vspace{-0.2cm}
    \setlength{\tabcolsep}{3pt}
    \begin{tabular}{{
    l
    S[table-format=1,round-precision=0]
    S[table-format=2.1,round-precision=1]
    S[table-format=2.1,round-precision=1]
    }}
    \toprule
    System & {\#CH} & {SDRi (dB)} & {SI-SDRi (dB)} \\
    \midrule
    SpEx+~\cite{ge2020spex+} & 1 & 10.0 & 10.9 \\
    X-TF-GridNet~\cite{hao2024x} & 1  & 14.2 & 15.3 \\
    X-CrossNet~\cite{kalkhorani2024tf} & 1 & 14.1 & 14.6 \\
    DCF-NEt~\cite{xue2025dualstream} & 1 & 14.5 & 15.8 \\
    USEF-TSE~\cite{zeng2025usef} & 1 & 14.9 & 16.1 \\
    CIENet-C2F-mDPTNet~\cite{yang2024coarse} & 1 & 16.0 & 17.5 \\
    LExt (TFGridNetV2)~\cite{shen2025listen} & 1 & 16.7 & 18.3 \\
    Multi-TasNet~\cite{zhang2020end} & 2 & {-} & 12.1 \\
    Conv-TasNet-CD$^{\ast}$~\cite{han2021multi} & 2 & 11.3 & 12.2 \\
    U-Conv-based Extraction~\cite{zhang2021time} & 2 & {-} & 13.4 \\
    HSRL-TSE$^{\ast}$~\cite{he2025enhancing} & 2 & 9.0 & 9.1 \\
    \midrule
    MC-LExt (TFGridNetV2) \& SI-SDR & 2 & \bfseries 18.5 & \bfseries 20.0 \\
    MC-LExt (TFGridNetV2) \& LOG-MSE & 2 & 18.2 & 19.7 \\
    \bottomrule
    \end{tabular}
    \label{tab:whamr}
\end{table}

\begin{figure}[!tp]
    \centering
    \vspace{-0.3cm}
    \includegraphics[width=1.0\linewidth]{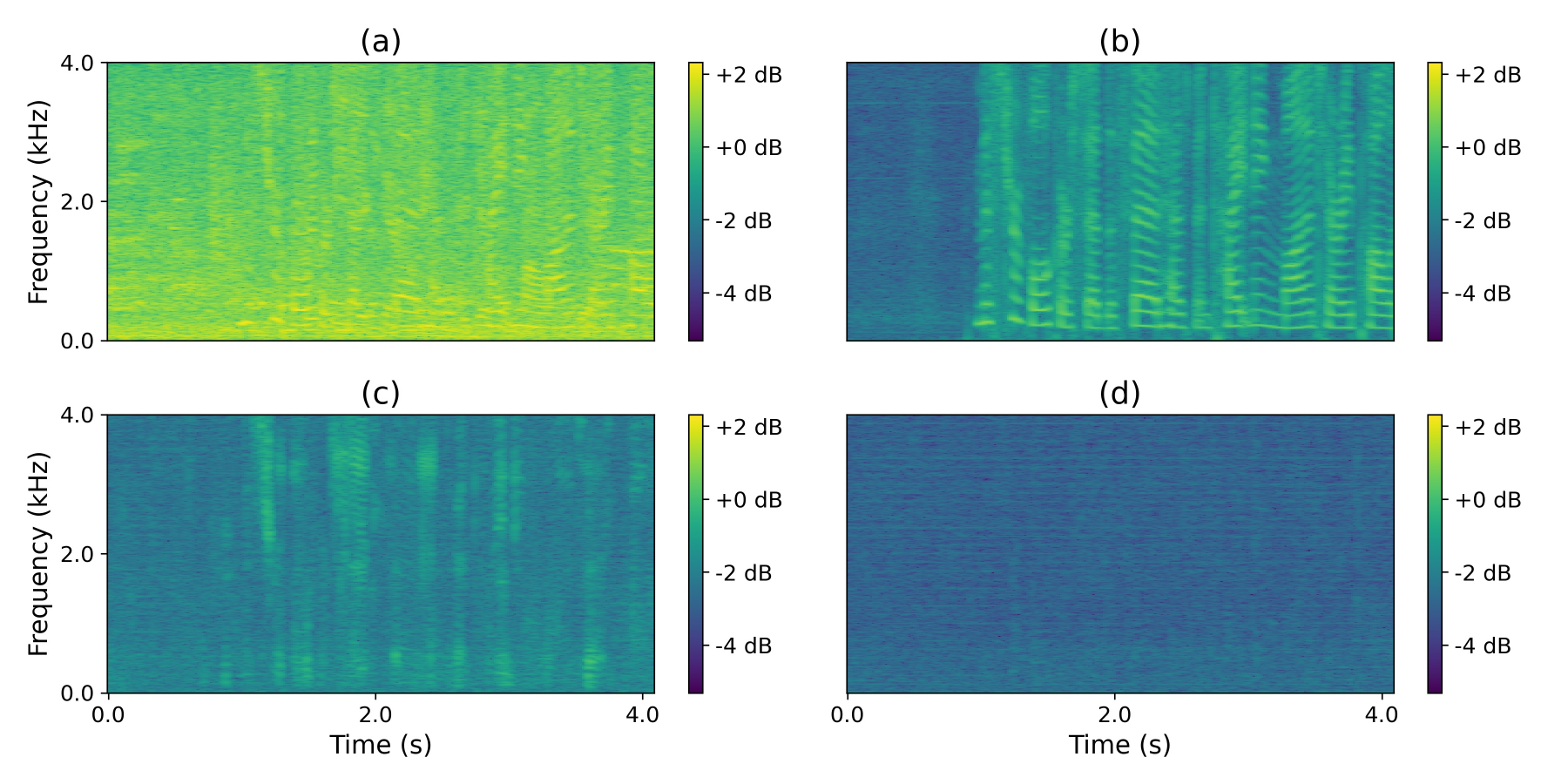}\vspace{-0.4cm}
    \caption{Illustration, based on a negative pair (the enrollment speaker is absent in the mixture), of log spectrograms of (a) mixture; (b) target speech; (c) estimated speech by Vanilla TSE; and (d) estimated speech by MC-LExt.}
    \vspace{-0.5cm}
    \label{fig:case}
\end{figure}

\section{Evaluation Results}\label{results}

Table~\ref{tab:ablation} compares MC-LExt with monaural LExt based on the WHAMR! dataset.
Compared to monaural LExt, MC-LExt achieves clearly better SDRi and SI-SDRi across all configurations, demonstrating the benefit of leveraging multi-channel information.
Compared to a \textbf{Vanilla TSE} baseline which conditions the DNN only on a speaker embedding, MC-LExt improves SI-SDRi from $17.4$ to $18.6$~dB, demonstrating its stronger speaker-aware modeling capability.
Enabling the downsampler (DS) reduces the enrollment embedding length, slightly improving performance (from $18.6$ to $18.8$ dB SI-SDRi) without harming model performance.
Meanwhile, enabling only the speaker embedding (SE) yields $18.9$ dB SI-SDRi, showing that conditioning the network on compact speaker identity features can strengthen TSE performance even without downsampling.
Combining DS and SE achieves the best results ($19.1$ dB SI-SDRi). 
We also experiment with the Log-MSE loss, which resulted in inferior performance ($18.8$ dB SI-SDRi), likely because Log-MSE focuses on signal energy rather than waveform structure, causing the model to be overly conservative on negative pairs and degrading reconstruction quality on positive pairs. 
In Table \ref{tab:energy}, we supply each mixture in WHAMR! test set with a speech signal uttered by a speaker different from the ones in the mixture, and report how silent the output is via an \textit{energy suppression ratio} metric defined as $10 \times \log_{10}(\|y\|_2^2 / \|\hat{s}\|_2^2)$.
From the results, we can see that MC-LExt can better produce silent outputs than Vanilla TSE.
In addition, for negative pairs, Fig.~\ref{fig:case} illustrates a case study showing that the vanilla TSE system trained via Log-MSE fails to remain silent, while MC-LExt successfully outputs silence.


Table~\ref{tab:fusion} investigates different fusion strategies for integrating speaker embeddings into MC-LExt. Multiplication achieves the highest performance, yielding $17.6$ dB SDRi and $19.1$ dB SI-SDRi.
FiLM-based conditioning performs competitively, slightly outperforming concatenation and addition. These results indicate that the choice of fusion mechanism can further boost the discriminative power of the speaker embedding in the extraction process.

Table~\ref{tab:forward_blocks} examines the trade-off between computational cost and performance by varying the number of forward blocks ($\mathcal{L}$) processing the enrollment segment. Reducing the number of enrollment forward blocks from $4$ to $1$ lowers the GMAC/S from $316.73$ to $198.48$ ($\approx$37\% reduction). Moreover, $\mathcal{L}=1$ and $\mathcal{L}=2$ produce very similar results (SI-SDRi difference $\textless$ $0.1$ dB), and $\mathcal{L}=3$ achieves the highest SI-SDRi ($19.2$ dB) and SDRi ($17.7$ dB). However, the performance trend is non-monotonic increasing and $\mathcal{L}=4$ yields the lowest scores despite having the largest computational cost. These results suggest that over-processing the enrollment segment may introduce redundancy or noise that slightly degrades extraction accuracy. We can conclude that forwarding fewer blocks for the enrollment is an effective strategy to reduce computational load without sacrificing extraction accuracy.
 
Table~\ref{tab:libri2mix} and \ref{tab:whamr} respectively compare the results of MC-LExt and state-of-the-art TSE systems on MC-Libri2Mix and WHAMR!.
On WHAMR!, MC-LExt significantly outperforms $1$- and $2$-channel baselines, including LExt~\cite{shen2025listen} and the U-Conv-based TSE model~\cite{zhang2021time}.
This highlights the effectiveness of our onset-prompting strategy combined with the downsampling and speaker embedding mechanisms for MC-TSE.
On the $4$-channel MC-Libri2Mix dataset, MC-LExt also delivers substantial improvements. Compared to a conventional spatial filtering based method (Mask-based MVDR)~\cite{erdogan2016improved} and recent DOA-based methods like L-SpEx~\cite{ge2022spex}, MC-LExt also delivers substantial improvements ($16.3$ dB SDRi and $14.7$ dB SI-SDRi). Compared with the strong L-SpEx baseline ($7.4$ dB SI-SDRi), MC-LExt more than doubles the improvement, demonstrating strong generalization to clean, multi-channel, and highly overlapped conditions.
These results confirm that MC-LExt not only excels in noisy-reverberant conditions but also maintains robustness and scalability across diverse datasets.

\section{Conclusions}\label{conclusion}


We have proposed MC-LExt, an onset-prompted framework for MC-TSE in noisy-reverberant environments, which compresses long enrollment speech while retaining the speaker embedding conditioning mechanism to integrate spatial and speaker identity cues.
Experiments on the WHAMR! and MC-Libri2Mix datasets show that MC-LExt consistently surpasses existing TSE models by a clear margin.

\bibliographystyle{IEEEtran}
{\footnotesize
\bibliography{ref.bib}
}

\end{document}